\documentclass[12pt,preprint,usegraphicx]{aastex}
\usepackage{emulateapj5}

\newcommand\msun{\ifmmode{\hbox{M$_\odot$}}\else{M$_\odot$}\fi}
\newcommand{\reduceme}{\mbox{R\raisebox{-0.35ex}{E}D%
\hspace{-0.05em}\raisebox{0.85ex}{uc}\hspace{-0.90em}%
\raisebox{-.35ex}{{m}}\hspace{0.05em}E}}

\slugcomment{}

\shorttitle{Differences in Ca indices between field and Coma Es}
\shortauthors{Cenarro et al.}

\begin{document}

\title{Early-Type Galaxies in the Coma Cluster: a new piece in the Calcium puzzle}

\author{Cenarro A. J.\altaffilmark{1}, S\'{a}nchez--Bl\'{a}zquez P., Cardiel N.\altaffilmark{2}, and Gorgas J.}
\affil{Departamento de Astrof\'{\i}sica, Facultad de F\'{\i}sicas, Universidad Complutense de Madrid, 28040 Madrid, Spain} 
\email{cen@astrax.fis.ucm.es}
\altaffiltext{1}{Also at UCO/Lick Observatory, University of California, Santa Cruz, CA 95064, USA}
\altaffiltext{2}{Also at Calar Alto Observatory, CAHA, Apdo. 511, 04044, Almer\'{\i}a, Spain}

\begin{abstract}
We present measurements of the Ca\,{\sc ii} triplet and the Ca4227
Lick-index for a sample of early-type galaxies in the Coma cluster,
deriving, for the first time, their corresponding relationships with
velocity dispersion. Compared with a similar subsample of ellipticals
in the field, Coma galaxies with velocity dispersions in the range
$\approx 180 - 270$\,km\,s$^{-1}$ exhibit significant differences in
the strengths of the Ca features, suggesting an influence of the
environment on the star formation histories of these galaxies. We
argue that the main scenarios previously proposed to explain the
relatively low Ca\,{\sc ii} triplet of galaxies are not able by
themselves to simultaneously reconcile the strengths of the two Ca
indices in both environments.
\end{abstract}


\section{Introduction}

The near-infrared (nIR) Ca\,{\sc ii} triplet constitutes the strongest
absorption feature in the integrated spectra around 8600\,\AA\ of
relatively old stellar populations. Given its well-known capability to
derive the metallicity of galactic globular clusters (Armandroff \&
Zinn~1988, hereafter A88; Rutledge, Hesser \& Stetson~1997), this
spectral feature was considered a useful tool to study the stellar
populations of normal galaxies.  However, the first observational work
(Cohen~1979; Bica \& Alloin~1987; Terlevich, D\'{\i}az \&
Terlevich~1990; Houdashelt~1995) reported only small differences among
the integrated Ca\,{\sc ii} triplet of several types of galaxies,
remarkably at variance with the trends found when using other
metallicity indicators (e.g. the Mg triplet at 5175\AA). The
difficulties of previous single stellar population (SSP) models to
make reliable predictions of the Ca\,{\sc ii} triplet (due to severe
deficiencies in the available stellar libraries) prevented an
understanding of the above apparent inconsistency.

After a series of papers devoted to construct realistic models for
disentangling the actual behaviour of the Ca\,{\sc ii} triplet in
stellar populations (Cenarro et al.~2001a, hereafter C01; Cenarro et
al.~2001b; Cenarro et al.~2002, hereafter C02; Vazdekis et al.~2003,
hereafter V03), the above topic is being revisited and constitutes a
controversial matter of debate. There exist recent evidences showing
that, unlike other metal-lines, the Ca\,{\sc ii} triplet is
surprisingly anti-correlated with central velocity dispersion
($\sigma_0$) for several types of galaxies: ellipticals (Es) in the
field (Saglia et al.~2002, hereafter S02; Cenarro et al.~2003,
hereafter C03), bulges of spirals (Falc\'on-Barroso et al.~2003,
hereafter F03) and dwarf Es (dEs) in the Fornax cluster (Michielsen et
al.~2003). Even more importantly, the measured values in giant Es lie
well below model predictions for any reasonable choice of stellar
populations parameters (age and metallicity). A wide range of
interpretations, like the existence of a dwarf-enriched stellar
population, Ca underabundances, Ca depletion in the interstellar
medium or a composite stellar population have been discussed in the
above references. Unfortunately, the lack of definitive evidences
makes no consensus to exist at the moment.

In the blue spectral range, Ca4227 is considered as the most
Ca-sensitive Lick-index (Tripicco \& Bell~1995). Intriguingly, despite
Ca is an $\alpha$-element like Mg, the metallicities inferred for Es
using Ca4227 suggest that Ca is not enhanced --even it could be
depressed-- with respect to Fe (Worthey~1992; Vazdekis et al.~1997,
hereafter V97; Worthey~1998; Trager et al.~1998; Vazdekis et al.~2001;
Proctor \& Sansom~2002, hereafter P02). Once again, the existence of
Ca underabundances have been demanded to explain the low Ca4227 values
(Thomas, Maraston \& Bender~2003b; hereafter T03b).

It is well known that the environment is expected to play a decisive
role in the assembling and star formation history of
galaxies. Hierarchical scenarios predict that Es in rich clusters were
mostly formed at high redshifts, whereas field Es may have
experienced an extended and more complex star formation history
(Kauffmann \& Charlot~1998). Although no differences in the [Mg/Fe]
ratios seem to exist between cluster and field Es (J{\o}rgensen~1999;
Kuntschner et al.~2002), S\'anchez-Bl\'azquez et al.~(2003, hereafter
S03) have recently found significant differences in C and N abundance
ratios between Coma and field Es. These differences impose strong
constraints to models of galaxy formation and chemical evolution.

The above picture motivated us to extend the previous work to the
analysis of the Ca\,{\sc ii} triplet and the Ca4227 index in galaxies
of different environments. First of all, because it might reveal new
clues on the so confusing Ca\,{\sc ii} triplet behaviour. Secondly,
because the fact that Ca seems to be an anomalous $\alpha$-element
makes it specially interesting from the point of view of chemical
evolution.

\section{Observations and data reduction}
\label{stellib}

This work makes use of long-slit spectroscopic data for two samples of
galaxies: Es from the Coma cluster and Es in the field. Both were
observed during two different runs (1999 March and 2001 April) using
ISIS, the double-arm spectrograph at the 4.2\,m William Herschel
Telescope (Observatorio del Roque de los Muchachos, La Palma). All
details about the sample, observations and data reduction for Es in
the field are given in C03 and S\'anchez-Bl\'azquez~(2004; hereafter
S04), for the nIR and blue-arm data respectively. We also refer the
reader to S04 for a description of the reduction of the blue
spectroscopic data for the Coma galaxies. In this section we just
concentrate on their nIR spectral range (8355 -- 9164\,\AA).

Our sample of galaxies from the Coma cluster consists of 28 early-type
galaxies (E -- S0) spanning a wide range of central velocity
dispersions ($80 \la \sigma_0 \la 370$\,km\,s$^{-1}$). We have
rejected the faintest low-mass galaxies in S03 because of quite large
errors. When it was feasible, the 3.5\,arcmin long slit (2\,arcsec
width, providing 2.9\,\AA\ FWHM spectral resolution) was rotated to
include two galaxies in the same exposure. Otherwise, single galaxies
were observed by aligning the slit with the major axis. Exposure times
of 1200 -- 2000\,s per galaxy leaded to signal-to-noise ratios per
angstrom from 30 to 120\,\AA$^{-1}$ in a central aperture of radius
$R_{\rm eff}$/8.

Standard spectroscopic reduction procedures were performed with
\reduceme\, (Cardiel~1999). Apart from an accurate correction for the
fringe pattern, we took special care on the sky subtraction and the
correction for telluric absorptions at $\lambda \ga 8950$\,\AA\ (by
using the telluric pattern derived from the spectra of flux standard
stars). Both are specially critical for Coma galaxies since the
redshift of the cluster (z $\sim$ 0.025) places some index bandpasses
in regions affected by atmospheric absorptions and strong sky emission
lines. The availability of error spectra for each galaxy frame allowed
us to estimate reliable uncertainties in the measurements of the
indices. Index errors account for photon statistics, uncertainties in
the flux calibration and radial velocity determinations (see details
in C01).

Final spectra were relative-flux calibrated using spectrophotometric
standard stars observed several times at different air masses. In
order to transform the nIR spectra to the V03 spectrophotometric
system, a sample of 39 stars (from B to late M spectral types) in
common with C01 were observed during twilights. Following C03, they
were also employed as templates for $\sigma_0$ determinations. Rather
than using template-dependent polynomials, systematic differences
between indices measured at different spectral resolutions were
prevented by broadening all the spectra up to the largest $\sigma_0$
of the galaxy sample, namely 370\,km\,s$^{-1}$, corresponding to
NGC\,4889.

The observation of four Es in common with C03 allowed us to confirm 
an homogeneous spectrophotometric system for the blue and nIR spectra
of both observing runs.

\section{Index--$\sigma_0$ relations}
\label{indexsigma}

We analyze, for the first time, the behaviours of the Ca\,{\sc ii}
triplet, the H Paschen series (CaT$^*$, CaT and PaT line-strength
indices; see definitions in C01) and the Ca4227 Lick-index as a
function of $\sigma_0$ for Es in the Coma cluster. All the
measurements correspond to a central aperture of radius $R_{\rm
eff}$/8 (or 1\,arcsec for galaxies with $R_{\rm eff} <
8$\,arcsec). There exists a database{\footnote {\tt
http://www.ucm.es/info/Astrof/ellipt/CATRIPLET.html}} listing the
galaxy sample as well as the indices and $\sigma_0$ determinations.
 
Figure\,1 shows the measurements of the above indices versus
$\sigma_0$ for Es in the Coma cluster and, for comparisons, for a
similar subsample of Es in the field. Error-weighted, least-squares
linear fits to all data have been computed for both samples (see
labels in Fig.\,1). CaT$^*$ and CaT in Coma Es exhibit an
anti-correlation with $\sigma_0$ (panels $a$ and $b$) which is
compatible with the one derived for field Es (C03). However, because
of the errors of the computed slopes, we cannot rule out that a flat
trend with $\sigma_0$ could exist for the Coma sample. As in the case
of field Es, a flat behaviour is derived for the PaT index (panel
$c$), whilst both slightly positive and negative (nearly flat) trends
are evidenced for the Ca4227 indices of field and Coma Es respectively
(panel $d$). In brief, as regards to the qualitative behaviours with
$\sigma_0$ of the above indices, we find no important differences
between field and Coma Es.

The most striking differences between both samples arise from the mean
absolute values of their Ca indices. In particular, though
intermediate-low mass Es ($\sigma_0 \la 180$\,km\,s$^{-1}$) in both
samples exhibit similar mean index values, we find clear differences
for those Es within the range $180 \la \sigma_0 \la 270$\,km\,s$^{-1}$
(in the following we will refer to them as {\it massive} Es, the ones
being the main subject of discussion in this letter). Note that, other
than visual inspection, there exist not any {\it a priori} argument to
justify the above subsample selection. Interestingly, the fact that
the differences in C4668 and CN$_2$ for Coma and field Es are also
particularly evident in a similar range of velocity dispersion (S03)
supports the idea that massive Es in different environments could
indeed have different stellar population properties. Unfortunatelly,
the scarcity of Es with $\sigma_0 \ga 270$\,km\,s$^{-1}$ in our Coma
sample prevents us to confirm whether the above behaviour extends to
the high-mass end of the Es family.

For this subsample of massive galaxies consisting of 17 Coma Es and 15
Es in the field, we have computed error-weighted means of the indices
and their corresponding standard errors (see labels in Fig.\,1). The
statistical significance of the differences between the mean indices
of both samples is studied by using a $t$-test (see Fig.\,1; for a
significance level of $\alpha = 0.05$, $t$ values larger than 1.96
indicate that a significant difference exists). Apart from the fact
that no significant differences are found for the PaT index (as it is
expected for relatively old stellar populations; see V03), we conclude
the existence of a intriguing behaviour: while field Es exhibit Ca4227
values significantly larger than Coma Es, the contrary occurs for the
CaT$^*$ and CaT indices. In order to check that the above results are
not driven by uncertainties in the correction for nIR telluric
absorptions, we also measured the Ca2 and Ca3 classical indices by A88
(Ca2$+$3 $\equiv$ Ca2 $+$ Ca3). The high sensitivities of these
indices to velocity dispersion and to the presence of Paschen lines
(see C01) are not important when, as it is the case, they are measured
on the spectra of old stellar populations at equal spectral
resolution. In turn, the major improvement lies in the fact that their
continuum sidebands are located in regions free from telluric
absorptions for the redshift of the cluster. The data in Fig.\,1
(panel $e$) prove that the trends and differences inferred from these
indices are compatible with the ones derived for CaT$^*$ and CaT.

\section{Discussion}
\label{discussion}

As it has been recently reported by S02, C03 and F03, single and
composite stellar populations models with solar abundance ratios
([$\alpha/$Fe]$= 0.0$) and a standard initial mass function (IMF)
cannot reproduce the low Ca\,{\sc ii} values of massive galaxies by
solely assuming age and metallicity variations. Among others, two
alternate scenarios have been proposed: i) the existence of Ca
underabundances and ii) the existence of a dwarf-enriched stellar
population. In the following paragraphs we explore in turn whether the
above scenarios are consistent enough to explain simultaneously the
discrepancy with the model predictions and the differences between
distinct environments.

i) Even though there exist no evolutionary synthesis models for the
Ca\,{\sc ii} triplet in which non-solar abundances ratios are
considered, the Ca underabundant hypothesis has been proposed as a
plausible scenario to explain the low CaT of massive galaxies (S02;
F03). Making use of stellar population models with variable element
abundance ratios for the Lick indices (Thomas, Maraston \&
Bender~2003a; hereafter T03a), T03b suggest the existence of a Ca
depletion, with respect to the rest of $\alpha-$elements, that
increases with galaxy mass. Under this scenario, and assuming that the
integrated Ca\,{\sc ii} triplet indeed traces the Ca abundance, the
relative differences in the CaT of massive Coma and field Es should be
naturally explained as differences in the levels of Ca
underabundances, being field Es ``more Ca underabundant'' than Coma
galaxies. However, there exist a strong contradiction in the fact that
while mean Ca\,{\sc ii} indices of massive Coma Es are larger than of
those in the field, the contrary occurs for Ca4227. Furthermore,
taking into account that Ca4227 is affected by C and N abundances in
the sense that the stronger CN the lower Ca4227 (Schiavon et al.~2002;
P02; T03a), and given that field Es exhibit CN overabundances larger
than those in Coma Es (S03), the relative differences in Ca4227 cannot
be explained by just CN effects either. In fact, if we consider such
an effect, the actual difference between the Ca4227 strengths of both
samples should be even larger than that in Fig.\,1 (panel $d$). In
summary, just the existence of Ca underabundances and/or CN effects is
unable to reconcile Ca4227 and CaT$^*$ at the same time.

ii) Since the Ca\,{\sc ii} triplet of giant stars is stronger than
that of dwarf stars, the existence of a dwarf-enriched stellar
population could account for the low values of massive Es. This
scenario can be parametrized by means of a variation of the IMF-slope
($\mu$; assuming a Salpeter-like IMF). The evolutionary synthesis
models by V03 predict that, for old SSPs with intermediate-to-high
metallicities, CaT$^*$ and CaT mainly depend on $\mu$ (they decrease
with the increasing $\mu$ because of a larger dwarf-to-giant stars
ratio). In the light of the above predictions, massive Es are
interpreted to have both metallicities and dwarf-to-giant ratios
larger than low mass Es (C03). 

In Figure\,2 we make use of the models by Vazdekis (1999; hereafter
V99) and V03 to compare Coma and field Es under this scenario. The
only way to reconcile H$\beta$, [MgFe]' (the one being independent of
typical $\alpha$/Fe ratios as defined by T03a) and CaT$^*$ at the same
time is invoking different ages, metallicities and dwarf-to-giant
ratios. In turn, the later also affects the derived ages and
metallicities because of the non-negligible dependendence of H$\beta$
to the dwarf-to-giant ratio. From panels $a$, $b$, $c$ and $d$, we
derive similar metallicities (above solar) for both subsamples, a mean
age for field Es lower ($\sim 8.9$\,Gyr) than that for Coma Es ($\sim
11.2$\,Gyr), and mean dwarf-to-giant ratios larger for the former
($\mu \simeq 2.8$) than for the latter ($\mu \simeq 1.8$). However,
when this scenario is simultaneously applied to Ca4227, the model
predictions fail to reproduce the locus of the galaxies (panel
$c$). Therefore, even though relative differences between the Ca
indices of Coma and field Es could be compatible with just variations
of the dwarf-to-giant ratio (the larger $\mu$, the lower CaT$^*$ and
the larger Ca4227; panels $c$, $d$, $e$ and $f$), this argument is not
enough to reconcile absolute values of CaT$^*$ and Ca4227 by itself.

At this point, one notices that the current status is not quite
optimistic. Could it be that both scenarios coexist driving an
apparent inconsistency? Even though the sensitivity of Ca4227 to
unusual dwarf-to-giant ratios is not negligible, age and metal
abundances effects are mainly expected to govern its behaviour. The
arrows in panels $e$ and $f$ of Fig.\,2 would indicate the offsets in
Ca4227 -arising not only from the existence of Ca subabundances but
also from possible dilution effects due to CN enhancements- required
to reproduce the SSPs properties in the previous panels. On the
contrary, several arguments support the idea that the Ca\,{\sc ii}
triplet is rather unsensitive to Ca abundances. First, the CaT index
does not correlate with stellar [Ca/Fe] abundance ratios (C02). Also,
for SSPs older than $\sim 3$~Gyr with intermediate-to-high
metallicities, the integrated Ca\,{\sc ii} triplet saturates (or even
decreases for dwarf-enriched SSPs) with the increasing metallicity
(V03). This is indeed a by-product of the increasing luminosity-weight
of late-M stellar types (with very weak Ca\,{\sc ii} lines and strong
TiO bands in their spectra) as the temperature of the isochrone
becomes colder. Therefore, just because of temperature effects, very
metal rich stellar populations with high Ca abundances are expected to
exhibit CaT values lower than if, for example, they were
solar-abundant (see more details in V03). In this sense, Ca\,{\sc ii}
triplet indices are expected to be sensitive to global metallicity
rather than to Ca abundances.

To conclude, we report, for the first time, systematic differences
between the integrated Ca\,{\sc ii} triplet and Ca4227 of Es in the
Coma cluster and in the field which are difficult to reconcile at the
same time. Different abundance ratios and/or dwarf-enriched stellar
populations could be conspiring to produce apparently inconsistent Ca
indices. As proposed in S03, these kind of differences are probably
driven by the effect of the environment on chemical evolution and
galaxy formation. Unfortunately, far from concluding any definitive
prove for distinct star formation histories in massive Es located at
different environments, the fact is that the current picture of
calcium in galaxies seems to go beyond the capabilities of the
available SSPs models. An extensive effort in the field of
nucleosynthesis and stellar atmospheres are demanded to understand how
different abundance ratios affect, not only line-strength indices, but
also the physics and temperature of isochrones. In the mean time, the
collection of more and high-quality data of Ca line-strengths is
crucial for trying to understand the controversial Ca puzzle.

\section*{Acknowledgments}
The authors are indebted to the referee S.\,M.\,Faber for useful comments
and suggestions. A.\,J.\,C.~acknowledges financial support from a UCM
Fundaci\'on del Amo Fellowship. The WHT is operated on the island of La
Palma by the Royal Greenwich Observatory at the Observatorio del Roque
de los Muchachos of the Instituto de Astrof\'{\i}sica de
Canarias. This work was supported by the Spanish research project
AYA2003-01840.


\clearpage

\begin{figure*}[!t]
\scalebox{0.9}{\includegraphics{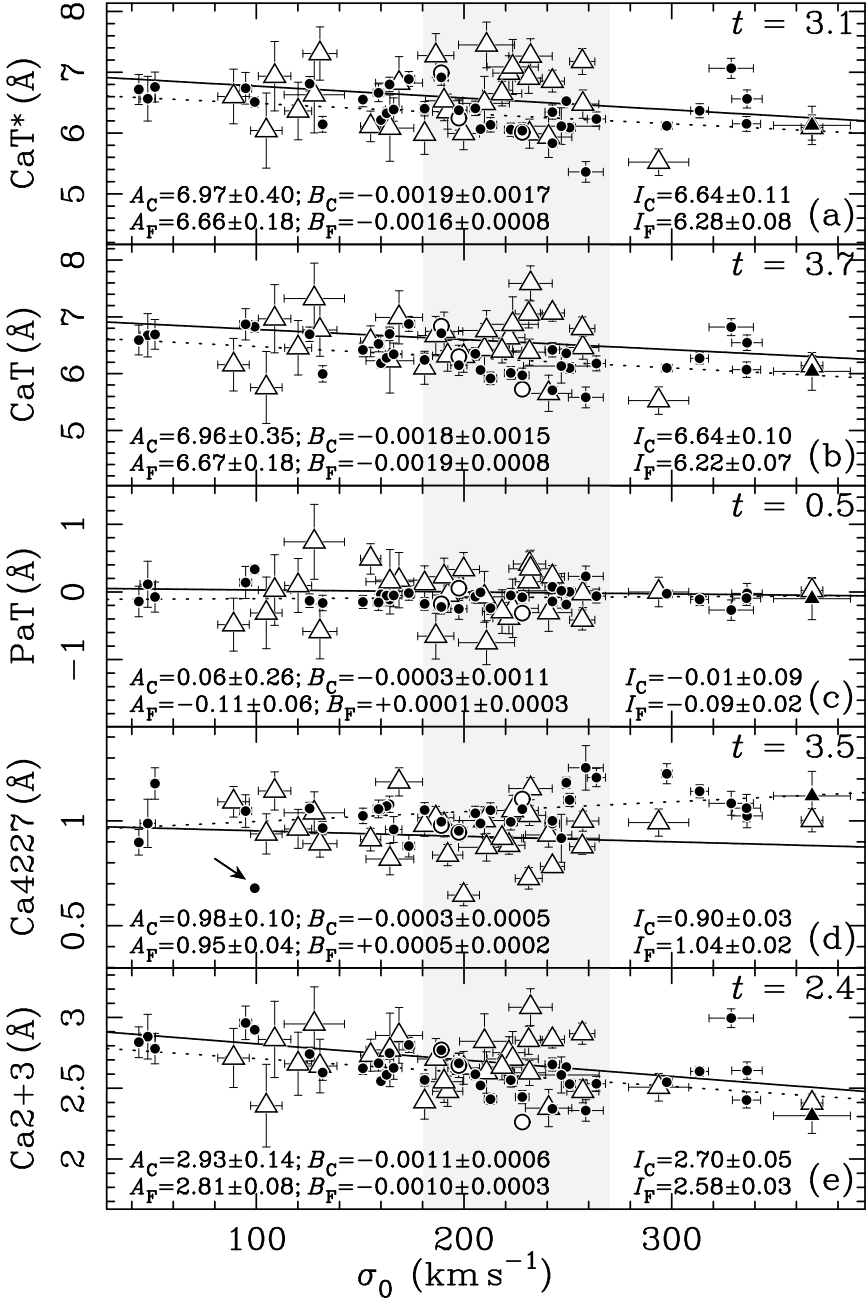}}
\caption{ CaT$^*$, CaT, PaT, Ca4227 and Ca2$+$3($\equiv$ Ca2 $+$ Ca3
indices by A88) versus central velocity dispersion for Es in the Coma
cluster (triangles) and in the field (circles). For comparisons, the
indices of 4 galaxies in common between both runs are included (open
circles and filled triangle). All the indices are measured at
$370$\,km\,s$^{-1}$ spectral resolution and corrected to the system
defined by the models (V03). Solid (Coma) and dotted (field) lines
represent error-weighted, least-squares linear fits to all data ($I =
A + B\,\sigma_0$; see the labels, with C and F subscripts refering to
Coma and field Es). $I_{\rm C}$ and $I_{\rm F}$ are error-weighted
mean indices for Coma and field Es within the range $180 \la \sigma_0
\la 270$\,km\,s$^{-1}$ (shaded region). A $t$ value larger than $\sim
1.96$ indicates that $I_{\rm C}$ and $I_{\rm F}$ are statistically
different. NGC\,4742 (see arrow in panel $d$), with a very low Ca4227
because of a central young stellar population (see C03), has been
rejected from the linear fits and the mean index computations.}
\label{fig1}
\end{figure*}

\begin{figure*}[!t]
\scalebox{1.50}{\includegraphics{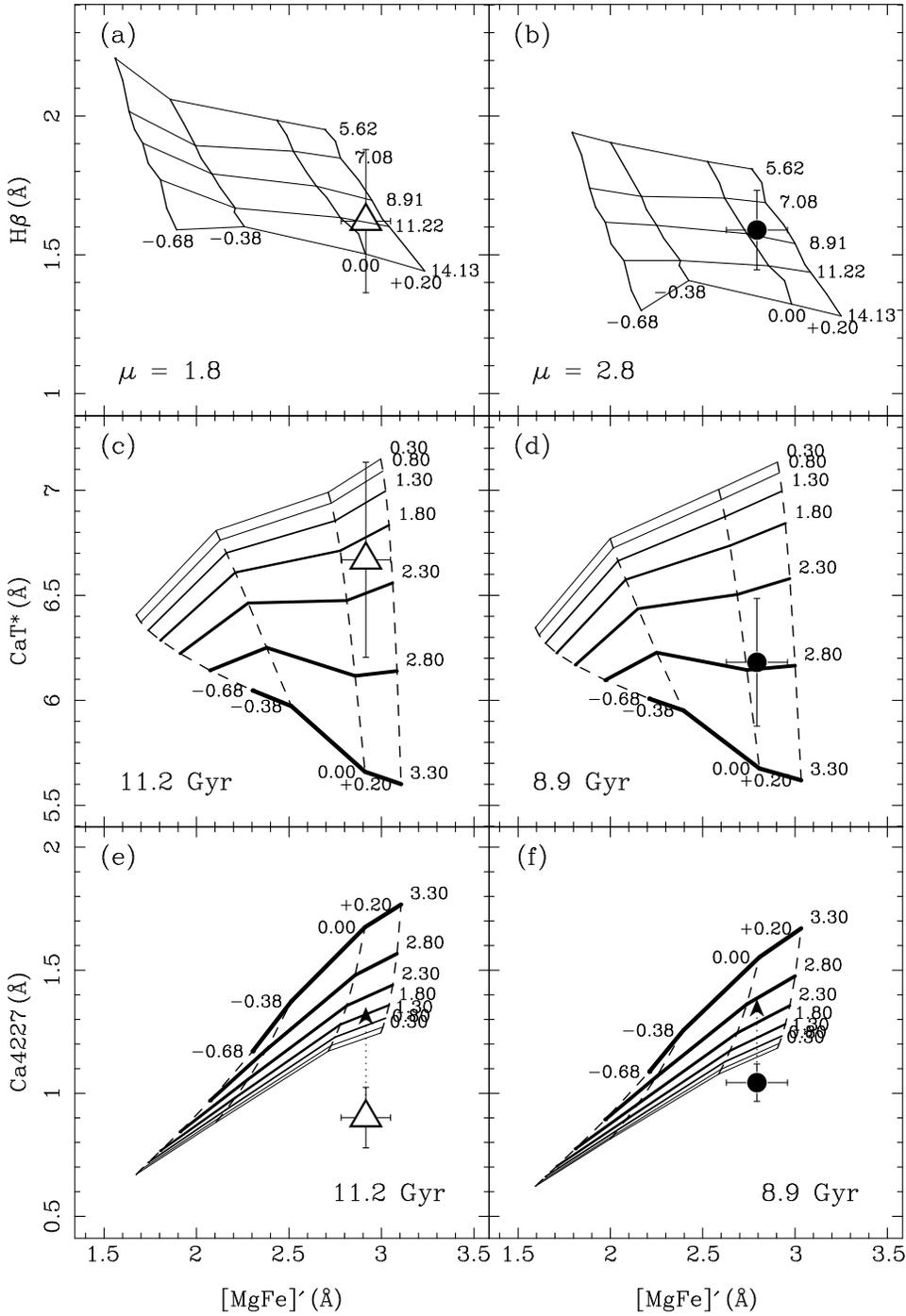}}
\caption{Mean values of H$\beta$, [MgFe]', Ca4227 and CaT$^*$ for Coma
(open triangle) and field (filled circle) Es within the range $180 \la
\sigma_0 \la 270$\,km\,s$^{-1}$. Rather than typical errors, error
bars correspond to the r.m.s.~standard deviation of the data (larger
than the former ones). SSPs model predictions by V03 and V99 at
$370$\,km\,s$^{-1}$ spectral resolution respectively are
overplotted. In panels $a$ and $b$, age and metallicity vary from
$5.62$ to $14.13$\,Gyr and from $-0.68$ to $+0.20$\,dex at different
Salpeter-like IMF slopes ($\mu$; see the labels). Panels $c$, $d$, $e$
and $f$ exhibit model predictions at fixed age (see the labels) with
varying $\mu$ ($0.3 - 3.3$) and metallicity (as in panels $a$ and
$b$). Arrows in panels $e$ and $f$ are explained in the text.}
\label{fig2}
\end{figure*}

\end{document}